\documentclass[12pt]{article}
\usepackage{graphicx}
\usepackage{amsmath}
\usepackage{mathrsfs}
\usepackage{mathrsfs}

\def\beq{\begin{equation}}
\def\eeq{\end{equation}}
\def\bea{\begin{eqnarray}}
\def\eea{\end{eqnarray}}
\def\nn{\nonumber}
\def\nl{\nonumber \\}
\def\roughly#1{\mathrel{\raise.3ex\hbox
{$#1$\kern-.75em\lower1ex\hbox{$\sim$}}}}
\def\lsim{\roughly<}
\def\gsim{\roughly>}
\def\sss{\scriptscriptstyle}

\def\bra#1{\left\langle  #1\right|} \def\ket#1{\left| #1\right\rangle}
\def\barpk{{\raise.35ex\hbox  {${\sss  (}$}}--{\raise.35ex\hbox{${\sss
)}$}}}        \def\bbarp{\hbox{$B$\kern-0.9em\raise1.4ex\hbox{\barpk}}}

\def\lsim{\roughly<}  \def\gsim{\roughly>}

\def\bra#1{\left\langle  #1\right|} \def\ket#1{\left| #1\right\rangle}

  \def\rr2{{1\over\sqrt{2}}}

\def\.{\!\cdot\!}    \def\:{\cdots}   \def\[{\left[}   \def\]{\right]}
\def\({\left(} \def\){\right)} 
\def\nn{\nonumber}
\def\nl{\nonumber \\}

\usepackage{epsfig}
\usepackage{dcolumn}
\usepackage{bm}

\newcommand{\comment}[1]{}
\newcommand{\newc}{\newcommand}

\def\lsim{\ ^<\llap{$_\sim$}\ }
\def\gsim{\ ^>\llap{$_\sim$}\ }
\def\r2{\sqrt 2}
\def\beq{\begin{equation}}
\def\eeq{\end{equation}}
\def\bea{\begin{eqnarray}}
\def\eea{\end{eqnarray}}
\def\nn{\nonumber}

\def\bra#1{\left\langle #1\right|}
\def\ket#1{\left| #1\right\rangle}

\def\lsim{\mathrel{\mathpalette\@versim<}}
\def\gsim{\mathrel{\mathpalette\@versim>}}
\def\@versim#1#2{\vcenter{\offinterlineskipresembel
    \ialign{$\m@th#1\hfil##\hfil$\crcr#2\crcr\sim\crcr } }}

\newc{\non}{\noindent}

\def\scat{ \nu_{\tau}+ n \to \tau^- + p}
\def\scatanti{ \bar{\nu}_{\tau}+ p \to \tau^+ + n}

\def\nutau{ \nu_{\tau}}
\def\numu{ \nu_{\mu}}

\def\taud{\tau^- \to \pi^- \nu_{\tau}}
\def\tauv{\tau^- \to \rho^- \nu_{\tau}}

\def\bra{\langle}
\def\ket{\rangle}

\begin{document}
 \unitlength = 1mm
\begin{flushright}
UMISS-HEP-2012-03 \\
[10mm]
\end{flushright}

\begin{center}
\bigskip {\Large  \bf Nonstandard interactions of tau neutrino via charged Higgs and $W'$ contribution}
\\[8mm]
Ahmed Rashed $^{\dag \ddag}$
\footnote{E-mail:
\texttt{amrashed@phy.olemiss.edu}}
, Murugeswaran Duraisamy $^{\dag}$ 
\footnote{E-mail:
\texttt{duraism@phy.olemiss.edu}} 
 and Alakabha Datta $^{\dag}$ 
\footnote{E-mail:
\texttt{datta@phy.olemiss.edu}} 
\\[3mm]
\end{center}

\begin{center}
~~~{\it $^{\dag}$ Department  of Physics and Astronomy,}\\ 
~~~{ \it University of Mississippi,}\\
~~~{\it  Lewis Hall, University, Mississippi, 38677 USA}\\
\end{center}

\begin{center}
~~~{\it $^{\ddag}$ Department  of Physics, Faculty of Science,}\\ 
~~~{\it  Ain Shams University, Cairo, 11566, Egypt}\\
\end{center}


\begin{center} 
\bigskip (\today) \vskip0.5cm {\Large Abstract\\} \vskip3truemm
\parbox[t]{\textwidth}  {We consider charged Higgs and $W'$ gauge boson contributions to the quasielastic scattering $\scat$ and $\scatanti$ . These effects  modify the standard model cross section for these processes and thus impact the extraction of the neutrino mixing angles $\theta_{23}$ and $\theta_{13}$ . We include form factor effects in our calculations and find the deviation of the actual mixing angle from the measured one, assuming the standard model cross section, can be significant and can depend on the energy of the neutrino. } 
\end{center}

\thispagestyle{empty} \newpage \setcounter{page}{1}
\baselineskip=14pt

\section{Introduction}
We now know that neutrinos have masses and that there is a leptonic mixing matrix just as there is a quark mixing matrix. This fact has been firmly established through a variety of solar, atmospheric, and terrestrial neutrino oscillation experiments. The next phase in the neutrino physics program is the precision measurement of the mixing angles, finding evidence for $CP$ violation  and measuring the absolute masses to resolve the mass hierarchy problem.

The existence of neutrino masses and mixing requires physics beyond the standard model (SM). Hence it is not unexpected that neutrinos could  have non-standard interactions (NSI). The effects of NSI have been widely considered in neutrino phenomenology
\cite{Wolfenstein:1977ue,Mikheev:1986gs,Roulet:1991sm,Brooijmans:1998py,
GonzalezGarcia:1998hj,Guzzo:1991hi,Bergmann:2000gp,Guzzo:2000kx,Guzzo:2001mi,Grossman:1995wx,Ota:2002na,Friedland:2005vy,Kitazawa:2006iq,Friedland:2006pi,Blennow:2007pu,EstebanPretel:2008qi,Blennow:2008ym,GonzalezGarcia:2001mp,Gago:2001xg,Huber:2001zw,Ota:2001pw,Campanelli:2002cc,Blennow:2005qj,Kopp:2007mi,Kopp:2007ne,Ribeiro:2007ud,Bandyopadhyay:2007kx,Ribeiro:2007jq,Kopp:2008ds,Malinsky:2008qn,Gago:2009ij,Palazzo:2009rb}.

It has been established that NSI cannot be an explanation for the standard oscillation phenomena,  but it may be present as a subleading effect. Many NSI involve flavor changing neutral current or charged current lepton flavor violating processes. In this paper we consider charged current interactions involving a charged Higgs and a $W'$ gauge boson in the quasielastic scattering processes $\scat$ and $\scatanti$.  In neutrino experiments, to measure the mixing angle the neutrino-nucleus interaction is assumed to be SM-like. If there is a charged Higgs or a $W'$ contribution to this interaction, then there will be an error in the extracted mixing angle. We will calculate the error in the extracted mixing angle. 

The  reaction $\scat$ is relevant for  experiments like 
Super-Kamiokande (Super-K) \cite{Abe:2012jj, Abe:2006fu} and OPERA \cite{:2011ph} that seek to measure $\nu_{\mu} \to \nu_{\tau}$ oscillation by the observation of the $\tau$ lepton . The above interaction is also important for the DONuT experiment \cite{Kodama:2007aa} which measured the charged-current (CC) interaction cross section of the tau neutrino. The DONuT central-value results for a $\nu_{\tau}$ scattering cross section  show  deviation from the standard model predictions by about  40\% but with large experimental errors; thus, the measurements are consistent with the standard model.  The new physics (NP) effects calculated in this paper modify the SM cross sections by less than 10\%
and are therefore consistent with the DONuT measurements. There have been recent measurements of the appearance of atmospheric  tau neutrinos by Super-K  \cite{Abe:2012jj} and by the OPERA Collaboration \cite{:2011ph}.

If high-energy Long Base Line (LBL) experiments (or atmospheric neutrino experiments scanning in the multi-GeV neutrino energy range) are designed to measure $\theta_{13}$ via $\nu_\tau$ ($\bar{\nu}_\tau$) appearance in the $\nu_{e} \to \nu_{\tau}$ ($\bar{\nu}_{e} \to \bar{\nu}_{\tau}$) oscillation mode, the scattering processes $\scat$ and $\scatanti$ will be important. 
The two  processes $\scat$ and $\scatanti$ involve the same new physics (NP) operators.

Generally, neutrino scattering contains contributions from various processes such as
quasielastic scattering (QE) , resonance scattering (RES), and deep inelastic scattering (DIS). Just above the threshold energy for $\tau$ production,  which is 3.45 GeV \cite{Abe:2012jj, Abe:2006fu}, the quasielastic interaction  dominates in  $\nu_\tau$ scattering \cite{Hagiwara:2003di, Conrad:2010mh}. At higher scattering energies other
processes have to be included. For instance,  
 the DIS is expected to be dominant above around 10 GeV \cite{Conrad:2010mh}, and so  $\nu_{\tau}$ scattering at the
 OPERA experiment, running at the average neutrino energy  $E_\nu = 17$ GeV \cite{:2011ph}, will be dominated by DIS. 
In this paper we will study the effects of NSI only in QE scattering, so we will limit ourselves to energies where QE is dominant. 
A full NSI analysis including all processes in  $\nu_{\tau}$ scattering will be discussed
in Ref.~\cite{datta_prep}.

There are several reasons to consider NSI involving the $( \nu_\tau, \tau)$ sector. First, the third generation may be more sensitive to new physics effects because of their larger masses. As an example, in certain versions of the two Higgs doublet models (2HDM) the couplings of the new Higgs bosons are proportional to the masses, and so new physics effects are more pronounced for the third generation. Second, the constraints on NP involving the third generation leptons are somewhat weaker, allowing for larger new physics effects. Interestingly, the branching ratio of $B$ decays to $\tau$ final states shows some tension with the SM  predictions \cite{belletau, Lees:2012xj} and this could indicate NP, possibly in the scalar or gauge boson sector \cite{Datta:2012qk}. Some examples of work that deals with NSI at the detector, though not necessarily involving the third family leptons, can be found in Refs.~\cite{nir, khalil, biggio}.

If there is NP involving the third generation leptons, one can search for it in $B$ decays such as $B \to \tau \nu_{\tau}$, $B \to D^{(*)} \tau \nutau$ \cite{nierste}, $ b \to s \tau^+ \tau^-$ etc. In general, the NP interaction in $B$ decays may not be related to the one in $\scat$ and $\scatanti$, and so these scattering processes probe different NP. The same NP in $\scat$ and $\scatanti$ can be probed in $\tau$ decays \cite{dattatau}, and we will consider the constraint on NP from this decay. However, in general, the scattering and the decay processes probe NP in different energy regions.

The form of NP in $\scat$ involves the operator $ {\cal{O}}_{NP}= \bar{u} \Gamma_i d \bar{\tau} \Gamma_j \nutau$, where $\Gamma_{i,j}$ are some Dirac structures. The process
$\scatanti$ gets a contribution from $ {\cal{O}}_{NP}^{\dagger}$. We will assume $CP$ conserving NP in this paper, and so the coefficients of the NP operators are real.
The same NP operator can also contribute to  hadronic tau decays $\taud$ and  $\tauv$, and the measured branching ratio of these decays can be used to constrain  the couplings in the operator ${\cal{O}}_{NP}$.
The ratio of the  charged Higgs contribution  to the SM in $\scat$ and $\scatanti $ is 
 roughly ${~(m_{N} / m_{\pi})}$ larger compared to the same ratio in $\taud$, where $m_{N, \pi}$ are the nucleon  and pion masses.
Hence, significant charged Higgs effects are possible in $\scat$ and $\scatanti$ even after imposing constraints from $\tau$ decays.
  We note that new interactions in the up and down quark sectors  can be constrained if one assumes  CKM unitarity. However, we do not consider this constraint as the NP in ${\cal{O}}_{NP}$ involves contributions from both the quark and the lepton sectors. 

As noted above, at the quark level NSI in $\scat$ and $\scatanti$ involve the $u$ and the $d$ quarks. Often in the analysis of NSI, hadronization effects of the quarks via  form factors are not included. As we show in our calculation the form factors play an important role in the energy dependence of the NP effects. In an accurate analysis one should also include nuclear physics effects which take into account the fact that the neutron and the proton are not free but bound in the nucleus. There is a certain amount of model dependence in this part of the analysis \cite{meloni}, and therefore we will not include nuclear effects in our calculation. Such effects can be easily incorporated once the free scattering cross sections are known.

The paper is organized in the following way. In the next section, we present a model-independent analysis of NP effects. In the following three sections, we consider the neutrino and antineutrino quasielastic scattering in the SM, in a model with charged Higgs and  in a model with an extra $W'$ gauge boson. In the last section, we present our conclusions.


\section{Model-independent analysis of new physics}

The process $\scat$ will impact the measurement of the oscillation probability for the $\numu \to \nutau$ transition and hence the extraction of the mixing angle $\theta_{23}$. The measurement of the atmospheric mixing angle $\theta_{23}$ relies on the following relationship \cite{relationship}:
\beq
N(\nu_\tau) = P(\nu_\mu \rightarrow \nu_\tau) \times \Phi (\nu_\mu)\times \sigma_{{\rm SM}}(\nu_\tau)\,,
\label{eq-1}
\eeq
where $N(\nu_\tau)$ is the number of observed events, $\Phi (\nu_\mu)$ is the flux of muon neutrinos at the detector,   $\sigma^{{\rm SM}}(\nu_\tau)$  is the total cross section of tau neutrino interactions with nucleons in the SM at the detector, and $P(\nu_\mu \rightarrow \nu_\tau)$ is the probability for the flavor transition $\numu \to \nutau$. This probability is a function of $(E,\; L,\; \Delta m_{ij}^2,\; \theta_{ij})$ with $i,j=1,2,3$, where $\Delta m_{ij}^2$ is the squared-mass difference, $\theta_{ij}$ is the mixing angle, $E$ is the energy of neutrinos, and $L$ is the distance traveled by neutrinos. The dominant term of the probability is
\beq
P(\nu_\mu \rightarrow \nu_\tau) \approx \sin^2 2\theta_{23} \cos^4 \theta_{13} \sin^2 (\Delta m^2_{23} L/4E).
\eeq
In the presence of NP, Eq.~\ref{eq-1} is modified as
\beq
N (\nu_\tau) = P(\nu_\mu \rightarrow \nu_\tau) \times \Phi (\nu_\mu)\times \sigma_{{\rm tot}}(\nu_\tau),
\label{eq-2}
\eeq
with $\sigma_{{\rm tot}}(\nu_\tau)=\sigma_{{\rm SM}}(\nu_\tau)+\sigma_{{\rm NP}}(\nu_\tau)$, where $\sigma_{{\rm NP}}(\nu_\tau)$ refers to the additional terms of the SM contribution towards the total cross section. Hence, $\sigma_{{\rm NP}}(\nu_\tau)$  includes  contributions from both the SM and NP interference amplitudes, and the pure NP amplitude. { }From Eqs.~(\ref{eq-1}, \ref{eq-2}), assuming $\theta_{13}$ to be small,{\footnote{The presence of NP impacts the extraction of the combination $\sin^2 2 \theta_{23}\cos^4 \theta_{13}$. The NP changes the extracted value of $\theta_{23}$ as well as $\theta_{13}$. But we fix the value of $\theta_{13}$ as an input at this point.}}
%
\bea
\label{modineq3}
\sin^2{2( \theta_{23})} &=& \sin^2{2( \theta_{23})_{SM}} \frac{1}{1 + r_{23}}\,,
\eea
where $\theta_{23}= (\theta_{23})_{SM} +\delta_{23}$ is the actual atmospheric mixing angle, whereas $(\theta_{23})_{SM}$ is the extracted mixing angle assuming the SM $\nu_\tau$ scattering cross section. Here $r_{23} = \sigma_{NP}(\nu_\tau) / \sigma_{SM}(\nu_\tau)$ is the ratio between the NP contribution, including the pure NP and interference terms, to the SM cross section. This means that $r_{23}$ can be positive or negative. Assuming negligible new physics effects in the $\mu-N$ interaction, the actual mixing angle $\theta_{23}$ is the same as the mixing angle extracted from the survival probability $P(\nu_\mu \rightarrow \nu_\mu)$ measurement. We will take the best-fit value for the mixing angle to be given by $\theta_{23} = 42.8^\circ$ \cite{GonzalezGarcia:2010er}. In other words, the presence of new physics in a $\nu_\tau$-nucleon scattering will result in the mixing angle, extracted from a $\nu_\tau$ appearance experiment, being different than the mixing angle from $\nu_\mu$ survival probability measurements. The relationship between the ratio $r_{23}$ and $\delta_{23}$ can be expressed in a model-independent form as
\bea
\label{modineq4}
 r_{23} &=&\Big[ \frac{\sin{2( \theta_{23})_{SM}}}{\sin{2( (\theta_{23})_{SM} + \delta_{23}) }}\Big]^2-1\,.
\eea

The probability of the tau antineutrino appearance $\bar{\nu}_e \to \bar{\nu}_\tau$ can be used to measure $\theta_{13}$ in the LBL experiments. In this case the effect of NP contributions to the process $\scatanti$ is pertinent. The relationship used in measuring $\theta_{13}$ will be given as
\beq
N(\bar{\nu}_\tau) = P(\bar{\nu}_e \rightarrow \bar{\nu}_\tau) \times \Phi (\bar{\nu}_e)\times \sigma_{{\rm tot}}(\bar{\nu}_\tau)\,,
\label{eq-111}
\eeq
where \cite{Donini:2002rm, Upadhyay:2011aj, Huber:2006wb}
\beq
P(\bar{\nu}_e \rightarrow \bar{\nu}_\tau) \approx \sin^2 2\theta_{13} \cos^2 \theta_{23} \sin^2 (\Delta m^2_{13} L/4E).
\eeq
Thus the relationship  between the ratio of the NP contribution to the SM cross section $ r_{13} = \sigma_{NP}(\bar{\nu}_\tau) / \sigma_{SM}(\bar{\nu}_\tau)$, which can be positive or negative, and $\delta_{13}$ can be obtained in a model-independent form as
\bea
\label{modineq444}
 r_{13} &=&\Big[ \frac{\sin{2( \theta_{13})_{SM}}}{\sin{2( (\theta_{13})_{SM} + \delta_{13}) }}\Big]^2-1\,.
\eea

\begin{figure}[htb]
\centering
\includegraphics[width=6.70cm]{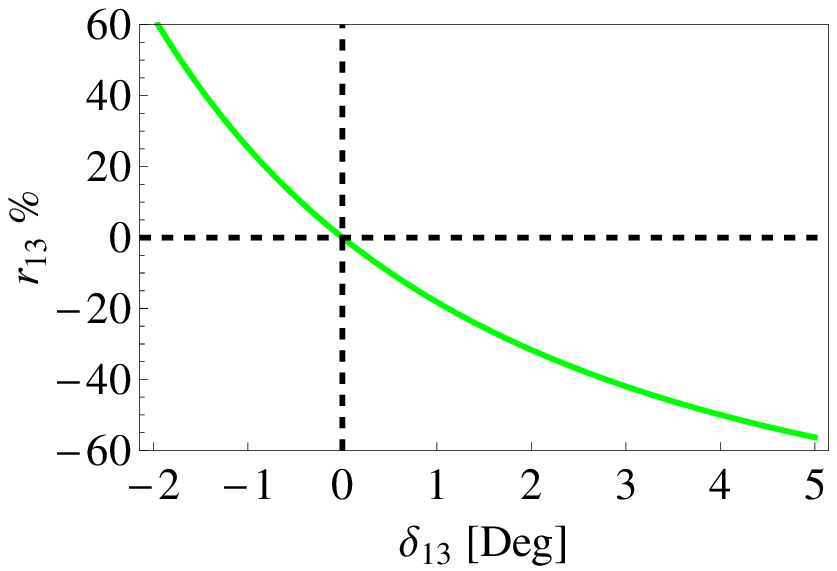}
\includegraphics[width=6.40cm]{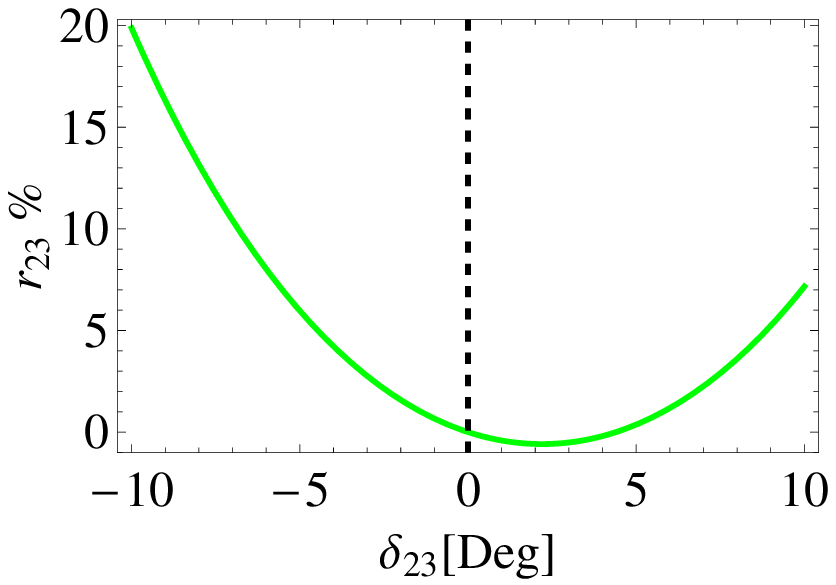}
\caption{Correlation plot for $r_{23} = \sigma_{NP}(\nu_\tau) / \sigma_{SM}(\nu_\tau) \%$ versus $\delta_{23} [Deg]$, and $r_{13} = \sigma_{NP}(\bar{\nu}_\tau) / \sigma_{SM}(\bar{\nu}_\tau) \%$ versus $\delta_{13} [Deg]$.}
\label{r23del23fig}
\end{figure} 

In Fig.~\ref{r23del23fig} we show the correlation between  $ r_{23(13)} \%$ and $\delta_{23(13)}$ [Deg]. One can see that  $\delta_{23}  \sim - 5^\circ $  requires $ r_{23} \sim 5\% $. But $\delta_{13}  \sim - 1^\circ $  requires $ r_{13} \sim 25\%$. In the following sections, we consider specific models of NP to calculate $r_{23}$ and $r_{13}$. We will consider a model with a charged Higgs and a  $W'$ model with both left- and right-handed couplings.


\section{Quasielastic neutrino scattering off free nucleon $-$ the SM}

In this section we consider the SM contribution to $\scat$ and $\scatanti$.
We first summarize  the SM results for the quasielastic scattering of a neutrino on a free  neutron target,
\beq
\label{quasisceq1}
\nu_l(k) + n(p) \to l^-(k^\prime) + p(p^\prime)\,,
\eeq
where $k$, $k^\prime$, $p$, and $p^\prime$ denote the four-momenta and  $l$ indicates the lepton $e, \mu$, or $\tau$.  The spin-averaged matrix element squared for the above reaction  is a convolution of spin-averaged leptonic and hadronic tensors  $L^{\mu \nu}$ and $H^{\mu \nu}$:
\beq
\label{ME}
|\bar{\mathcal{M}}|^2 = \frac{G^2_F}{2} L^{\mu \nu} H_{\mu \nu}.
\eeq
The leptonic tensor calculation is straightforward, but the  hadronic tensor involves nonperturbative effects. In order to calculate the hadronic tensor, we define the charged hadronic current for this process:
  \bea
 \label{hadcureq}
\bra p(p^\prime)|J^+_\mu |n(p) \ket &=&  V_{ud}\; \bra p(p^\prime)|(V_\mu -A_\mu)| n(p) \ket \nonumber\\
&=& V_{ud}\; \bar{p}(p^\prime) \Gamma_\mu n(p).
 \eea
The  expressions for the matrix elements of the vector and axial-vector currents  are summarized in terms of six form factors in the Appendix. Due to time reversal invariance, the form factors are real functions of $t = q^2$. When invariance under charge conjugation holds, two form factors vanish ($F_S = 0,\; F_T =0 $) \cite{SM-calc}. The matrix element, then, can be written as
\bea
\mathcal{M} &=& \frac{G_F \cos \theta_{c}}{\sqrt{2}}\bar{u_l}(k')\gamma^\mu (1-\gamma_5)u_{\nu_l}(k)\nonumber\\
 &&\bar{u}_{N'}(p') \left[  F^V_1 (t) \gamma_\mu + F^V_2 (t) i \frac{\sigma_{\mu\nu} q^\nu}{2M} + F_A (t) \gamma_\mu \gamma_5 + F_P (t) \gamma_5 \frac{q_\mu}{M} \right] u_{N}(p), 
 \label{matrixelementSM}
\eea
where $ N$ and $N'$ are the initial and final nucleons, while $l$ and $\nu_l$ are the final charged
lepton and the initial neutrino. In our case $N = n$, $N^\prime = p $, $ l = \tau $, and $\nu_l = \nu_\tau$.

After evaluating $|\bar{\mathcal{M}}|^2$, one can  obtain the  SM differential cross section for  the reaction in Eq.~(\ref{quasisceq1}) \cite{SM-calc},
\bea
\label{SMdiff}
 \frac{d\sigma_{SM}(\nu_l)}{dt} &=& \frac{M^2 G^2_F \cos^2{\theta_c}}{8 \pi E^2_\nu } \Big[A_{SM} + B_{SM} \frac{(s-u)}{M^2}  + C_{SM} \frac{(s-u)^2}{M^4}   \Big] \,,
\eea
where $G_F = 1.116637\times 10^{-5}$ GeV$^{-2}$ is the Fermi coupling constant, $\cos{\theta_c} = 0.9746 $ is the cosine of the Cabibbo angle, $M_W$ is the $W$ boson mass, and $E_{\nu}$ is the incident neutrino energy. $M = (M_p + M_n)/2 \approx 938.9$ MeV is the nucleon mass, and we neglect the proton-neutron mass difference. The expressions for the coefficients $f_{SM}\; (f = A, B, C)$  are  summarized in the Appendix. The Mandelstam variables are defined by  $s = (k+p)^2$, $t = q^2= (k-k^\prime)^2$, and $ u = (k-p^\prime)^2$.  The expressions for these variables in terms of  $E_\nu$ and the lepton energy $E_l$ are given in the Appendix.

The quasielastic scattering of an antineutrino on a free nucleon is given by 
\beq
\label{quasisceq111}
\bar{\nu}_l(k) + p(p) \to l^+(k^\prime) + n(p^\prime)\,.
\eeq
The charged hadronic current becomes \cite{Hagiwara:2003di, Llewellyn Smith:1971zm}
  \bea
 \label{hadcureq000}
\bra n(p^\prime)|J^-_\mu |p(p) \ket &=& \bra p(p)|J^+_\mu |n(p^\prime) \ket^\dagger \nonumber\\
&=&  V_{ud}\; \bar{n}(p^\prime) \tilde{\Gamma}_\mu p(p),
 \eea
where
\beq
\tilde{\Gamma}_\mu (p,p')= \gamma_0 \Gamma_\mu^\dagger (p',p)\gamma_0 .
\eeq
The relationship between the differential cross sections of $\scat$ and $\scatanti$ is \cite{Llewellyn Smith:1971zm, Strumia:2003zx}
\beq
 \frac{d\sigma_{SM}(\nu_l)}{dt}(s,t,u) = \frac{d\sigma_{SM}(\bar{\nu}_l)}{dt} (u,t,s).
\eeq
Thus, the matrix element is given by Eq.~\ref{matrixelementSM}, and the differential cross section, similarly to Eq.~\ref{SMdiff}, is given by 
\bea
\label{SMdiff000}
 \frac{d\sigma_{SM}(\bar{\nu}_l)}{dt} &=& \frac{M^2 G^2_F \cos^2{\theta_c}}{8 \pi E^2_\nu } \Big[A_{SM} - B_{SM} \frac{(s-u)}{M^2}  + C_{SM} \frac{(s-u)^2}{M^4}   \Big] \,.
\eea
The negative sign of $B_{SM}$ leads to a relatively smaller cross section for the antineutrino scattering.


\section{Quasielastic neutrino scattering off free nucleon $-$ Charged Higgs Effect}


We consider here the charged Higgs contribution to $\scat$ and $\scatanti $. Charged Higgs particles appear in multi-Higgs models. In the SM the Higgs couples to the fermion masses, but in a general multi-Higgs model the charged Higgs may not couple to the mass.
What is true in most models is that the coupling of the charged Higgs to the leptons is no longer universal. Hence, the extraction of $\theta_{23}$ and $\theta_{13}$ from $ \numu \to \numu$ and $ \bar{\nu}_e \to \bar{\nu}_e$ survival probabilities, respectively, will be different from $ \numu \to \nutau$ and $ \bar{\nu}_e \to \bar{\nu}_\tau$ probabilities, respectively, in the presence of a charged Higgs effect. 

The most general coupling of the charged Higgs is
\bea
\label{HiggsLag}
\mathcal{L} &=&\frac{g}{2\sqrt{2}}\left[ V_{u_i d_j} \bar{u}_i( g^{u_i d_j}_S \pm g^{u_i d_j}_P \gamma^5) d_j +  \bar{\nu}_i (g^{\nu_i l_j}_S \pm g^{\nu_i l_j}_P \gamma^5) l_j \right] \; H^{\pm}  ,
\eea
where $u_i$ and $d_j$ refer to up and down type quarks, and  $\nu_i$ and $l_j$ refer to neutrinos and charged leptons. The other parameters are as follows: $g = e/\sin{\theta_W}$ is the SM weak coupling constant, $V_{u_i d_j}$ is the CKM matrix element, and $g_{S,P}$ are the scalar  and pseudoscalar couplings of the charged Higgs to fermions. Here, in this work, we assume the couplings $g_{S,P}$ are real.

We will choose the couplings $g_{S,P}$, relevant for $\scat$ and $\scatanti$, to be given by  the two Higgs doublet model of type II (2HDM II). 
In the 2HDM II these couplings are related to couplings in other sectors and so can be constrained by measurements in these other sectors.
However, in our analysis, to keep things  general we will not assume any relation between the couplings $g_{S,P}$  and the couplings in other sectors, thereby avoiding constraints from other sectors.
To constrain the couplings $g_{S,P}$ we will only consider processes that are generated by ${\cal{O}}_{NP}=\bar{u} \Gamma_i d \bar{\tau} \Gamma_j \nutau$. In the 2HDM II,  constrains on the model parameters  come from various sectors \cite{2HDMII}. These constraints turn out to be similar but slightly stronger than the ones obtained in  our analysis.

The  coupling of charged Higgs boson ($H^{\pm}$) interactions to a SM fermion in the 2HDM II is \cite{Diaz:2002tp}
\beq
\label{HiggsLag1}
\mathcal{L} = \frac{g}{\sqrt{2}M_W} \sum_{ij} \Big[m_{u_i}  \cot{\beta} ~  \bar{u}_{i} V_{ij} P_{L,R} d_{j} + m_{d_j} \tan{\beta} ~  \bar{u}_{i} V_{ij} P_{R,L} d_{j}  + m_{l_j} \tan{\beta}~ \bar{\nu}_{i} P_{R,L} l_{j} \Big] H^{\pm},
\eeq
%
 where $P_{L,R}= {( 1 \mp \gamma^5) / 2}$, and $\tan \beta$ is the ratio between the two vacuum expectation values (vev's) of the two Higgs doublets. Comparing Eq.~(\ref{HiggsLag}) and Eq.~(\ref{HiggsLag1}), one can obtain
\bea
\label{2HDMcoup}
g^{u_i d_j}_S &=&  \left (\frac{m_{d_j} \tan{\beta} + m_{u_i} \cot{\beta}}{M_W} \right), \nonumber\\
g^{u_i d_j}_P &=&  \left (\frac{m_{d_j} \tan{\beta} - m_{u_i} \cot{\beta}}{M_W} \right),\nonumber\\
g^{\nu_i l_j}_S &=& g^{\nu_i l_j}_P = \frac{m_{l_j} \tan{\beta}}{M_W}.
\eea

\begin{figure}[htb!]
\centering
 \includegraphics[width=7cm]{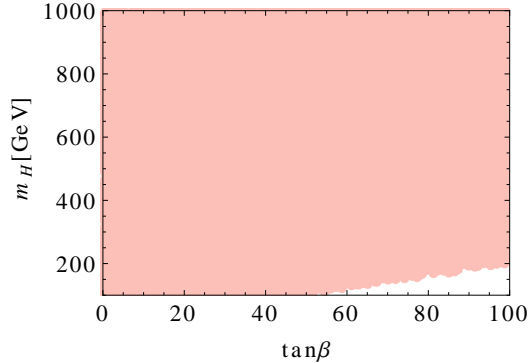}~~~
\caption{Constraint by $Br(\tau^- \to \pi^- \nu_\tau)$ at 95 \% CL. The colored region is allowed. }
\label{taupiplot}
\end{figure}

Constraints on the size of the operator ${\cal{O}}_{NP}= \bar{u} \Gamma_i d \bar{\tau} \Gamma_j \nutau$ can be obtained from the branching ratio  of the decay $\taud$. In the presence of a charged Higgs, the branching ratio for this  process is
\begin{eqnarray}
Br^{SM+ H}_{\tau^- \to \pi^- \nu_\tau} &=&  Br^{SM}_{\tau^- \to \pi^- \nu_\tau}  (1+ r^{\pi^2}_H)\,,
\end{eqnarray}
where the charged Higgs contribution is
\begin{eqnarray}
r^\pi_H &=&  \Big(\frac{m_u -m_d \tan^2{\beta}}{m_u + m_d} \Big) \frac{m^2_\pi}{m^2_H}\, .
\end{eqnarray}
The SM  branching ratio is related to the tau lepton width $(\Gamma_\tau)$ and the  decay rate($\Gamma^{SM}_{\tau^- \to \pi^- \nu_\tau}$) as $ Br^{SM}_{\tau^- \to \pi^- \nu_\tau} = 
\Gamma^{SM}_{\tau^- \to \pi^- \nu_\tau}/\Gamma_\tau$ with 
\begin{eqnarray}
\Gamma^{SM}_{\tau^- \to \pi^- \nu_\tau} &=&  \frac{G^2_F}{16 \pi} |V_{ud}|^2 f^2_\pi m^3_\tau \Big(1-\frac{m^2_\pi}{m^2_\tau}\Big)^2 \delta_{\tau/\pi}\,.
\end{eqnarray}
Here $\delta_{\tau/\pi} =1.0016 \pm 0.0014$ \cite{rad} is  the radiative correction. Further, the SM  branching ratio can also be expressed as 
\cite{barish}
\begin{eqnarray}
Br^{SM}_{\tau^- \to \pi^- \nu_\tau}  &=&  0.607 Br(\tau^- \to \nu_\tau e^- \bar{\nu}_e) = 10.82 \pm 0.02 \% \,,
\end{eqnarray}
while the  measured $Br(\tau^- \to \pi^- \nu_\tau)_{exp} = (10.91 \pm 0.07)\%$ \cite{pdg}.
In Fig.~\ref{taupiplot} we show the constraints on $m_H-\tan\beta$ from $\taud$. 
{} From Eq.~\ref{HiggsLag1} we can construct the NSI parameters  defined in Ref~\cite{biggio}  as $\varepsilon_{\tau\tau}^{ud(L)} \equiv \frac{m_u m_\tau}{m_H^2}$ and $\varepsilon_{\tau\tau}^{ud(R)} \equiv \frac{m_d m_\tau \tan^2 \beta}{m_H^2}$ . We find that the constraints on the effective operator considered in this work are consistent with the one in Ref. \cite{biggio}.
Finally, we note that $\tau$ has a significant branching ratio to $ \tau^- \to \rho^- \nu_{\tau}$ \cite{pdg}. However, a charged Higgs cannot contribute to this decay, and hence there is no constraint on the charged Higgs couplings from this decay \cite{dattatau}.

\begin{figure}[htb!]
\centering
 \includegraphics[width=5cm]{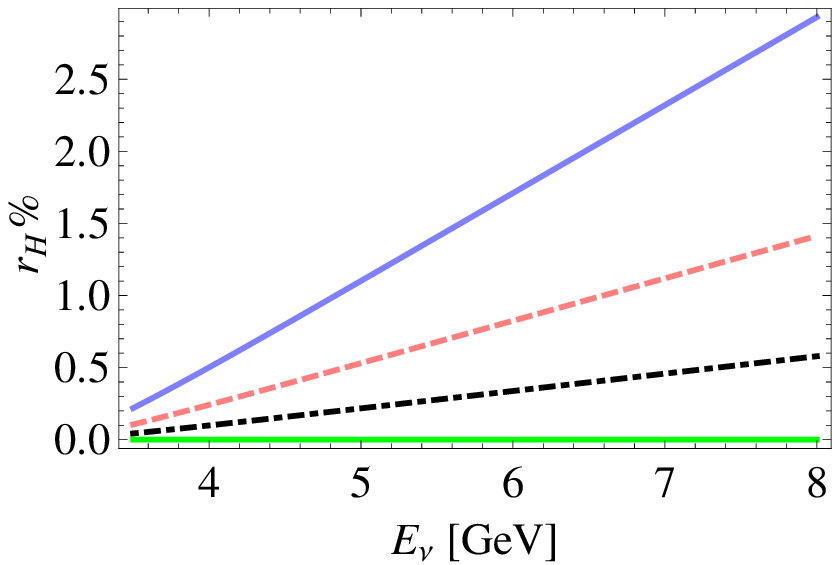}~~~
 \includegraphics[width=5cm]{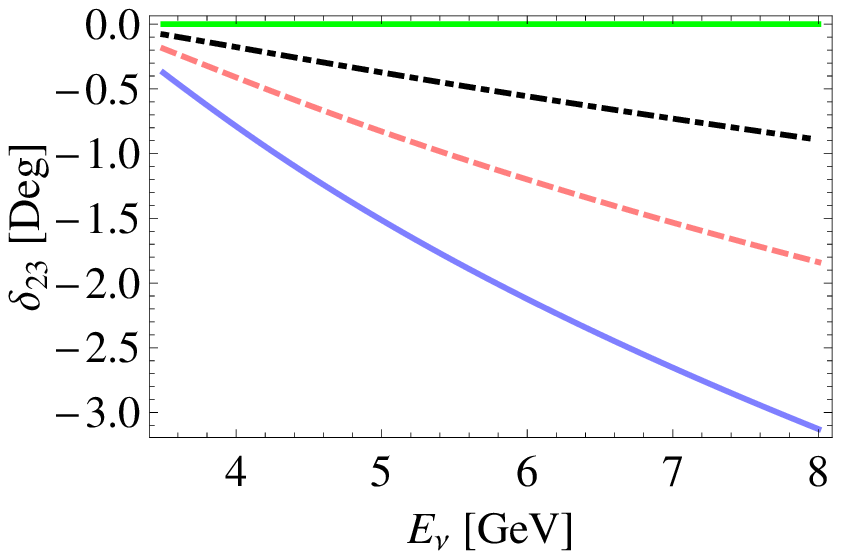}~~~
 \includegraphics[width=5cm]{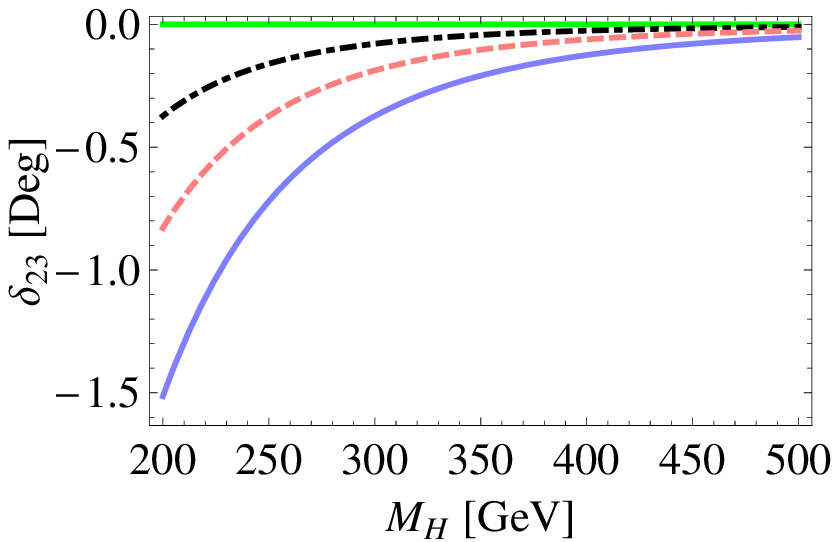}~~~
\caption{ Variation of $r_H^{23} \%$ with $E_\nu$ and variation of $\delta_{23}$ with $M_H$  and $E_\nu$. The green line corresponds to the SM prediction.  The black (dotdashed), pink (dashed), and blue (solid) lines correspond to $\tan{\beta} = 40, 50, 60$. The right figure is evaluated  at $E_\nu = 5$ GeV, while the left figures are evaluated at $M_{H} = 200$ GeV. Here, we use the best-fit value $ \theta_{23} = 42.8^\circ$ \cite{GonzalezGarcia:2010er}.}
\label{delHvsMHplot}
\end{figure}

Keeping in mind the constraints from Fig.~(\ref{taupiplot}), we calculate the charged Higgs contribution to $\scat$. The modified differential cross section for the reaction in Eq.~(\ref{quasisceq1}) is
\bea
\label{Higgsdiff}
\frac{d\sigma_{SM + H}}{dt} = \frac{M^2 G^2_F \cos^2{\theta_c}}{8 \pi E^2_\nu }  \Big[A_H +B_H \frac{(s-u)}{M^2}  +C_{SM} \frac{(s-u)^2}{M^4}  \Big],
\eea
where $x_{H} = m^2_W/M^2_{H} $, $A_H = A_{SM} + 2 x_{H} Re{(A^I_{H})} + x^2_{H} A^{P}_{H}$, and  $B_H = B_{SM} + 2 x_{H} Re{(B^I_{H})}$. Superscripts $I$ and $P$ denote the SM-Higgs interference  and pure Higgs contributions, respectively.  The expressions for the quantities $A^{I,P}_{H}$  and $B^{I}_{H}$ are given in the Appendix. The terms $A^{I}_{H}$  and $B^{I}_{H}$ are proportional to  the tiny neutrino mass, and we will ignore them in our calculation. Note that  this happens because we have chosen the couplings to be given by the 2HDM II. With general couplings of the charged Higgs, these interference terms  will be present.
The charged Higgs contribution relative to the SM $r_{H}^{23} = \frac{\sigma_H (\nu_\tau)}{\sigma_{SM}(\nu_\tau)}$ is proportional to $t$ because of the dominant term $x_t G^2_P$, where $x_t = t/4 M^2$ (see the Appendix for more details). Consequently, $r_{H}^{23}$ is proportional to the incident neutrino energy (see Fig.~(\ref{delHvsMHplot})). 
The deviation $\delta_{23}$ is  negative, as there is no interference with the SM;  hence, the cross section for $\scat$ is always larger than the SM cross section. This means that, if the actual $\theta_{23}$ is close to maximal, then experiments should measure $\theta_{23}$ larger than the maximal value in the presence of a charged Higgs contribution. 

\begin{figure}[htb!]
\centering
 \includegraphics[width=5cm]{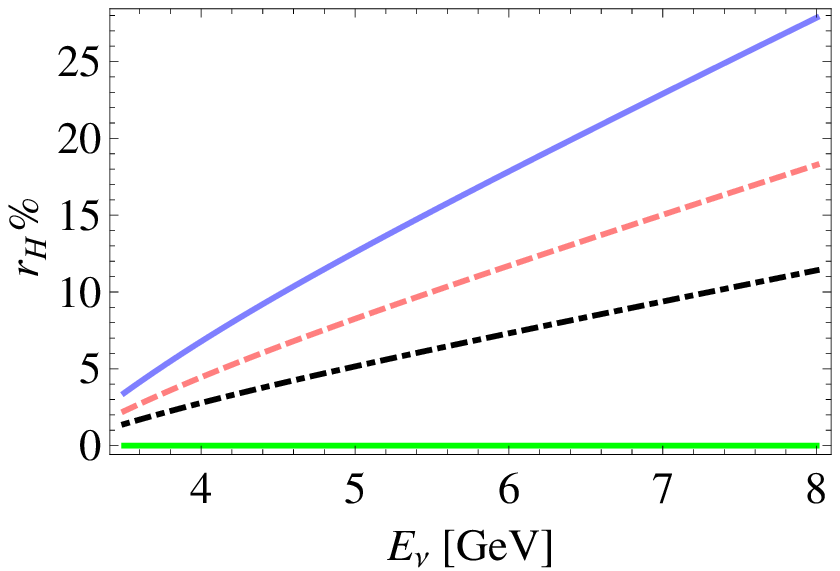}~~~
 \includegraphics[width=5cm]{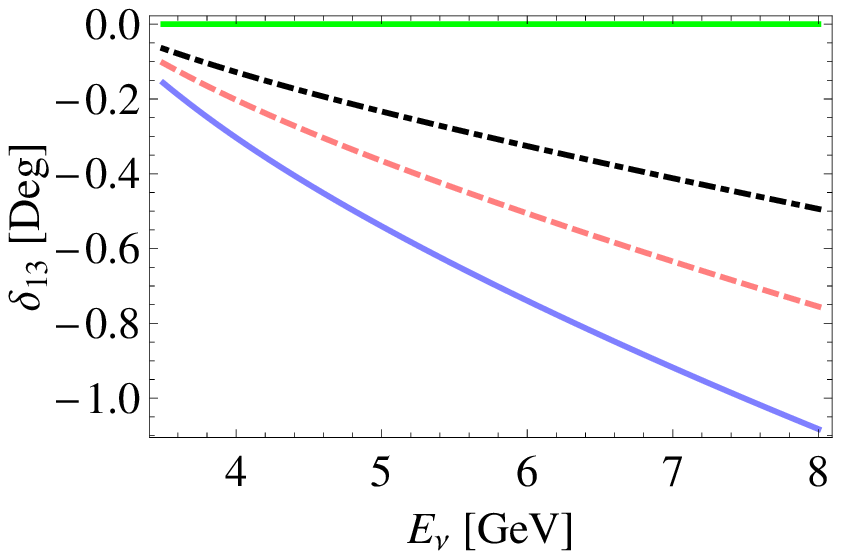}~~~
 \includegraphics[width=5cm]{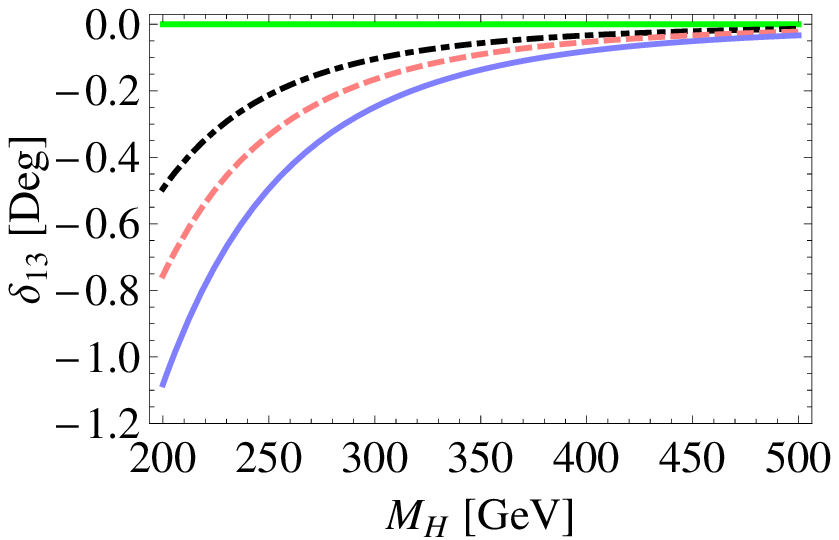}~~~
\caption{ Variation of $r_H^{13} \%$ with $E_\nu$ and the variation of $\delta_{13}$ with $M_H$  and $E_\nu$. The green line corresponds to the SM prediction.  The black (dotdashed), pink (dashed), and blue (solid) lines correspond to $\tan{\beta} = 80, 90, 100$. The right figure is evaluated  at $E_\nu = 8$ GeV, while the left figures are evaluated at $M_{H} = 200$ GeV. Here, we use the inverted hierarchy value $ \theta_{13} = 9.1^\circ$ \cite{Tortola:2012te}.}
\label{delHvsMHplot000}
\end{figure}

The differential cross section for the interaction $\scatanti$ has the same form as Eq.~\ref{Higgsdiff} in the limit of a massless neutrino.  The hadronic current  in this case is the complex conjugate of the one in the Appendix. The ratio $r_{H}^{13} = \frac{\sigma_H (\bar{\nu}_\tau)}{\sigma_{SM}(\bar{\nu}_\tau)}$, as well as the deviation $\delta_{13}$, is shown in Fig.~\ref{delHvsMHplot000}. As $\theta_{13}$ is a small angle, large $\tan \beta$ and small charged Higgs mass are preferred to produce an observable deviation $\delta_{13}$. {}For instance, we find $\delta_{13} \approx 1^\circ$ and $r_H^{13} \approx 30 \%$ at $E_\nu = 8$ GeV, $M_H = 200$ GeV, and $\tan \beta = 100$.

\section{Quasielastic neutrino scattering off free nucleon - $W^\prime$ model}

Many extensions of the SM contain a $W'$ gauge boson. We next consider modification to $\scat$ and $\scatanti$ in models with a $W^\prime$. There are limits on the $W^\prime$ mass from direct searches to
final states involving an electron and muon assuming SM couplings  for the $W^\prime$ \cite{pdg}. 
These limits generally do not apply to  the $W^\prime$ coupling to
$\nutau$ and $\tau$ which is relevant for our calculation.

The lowest dimension effective Lagrangian of $W^\prime$ interactions to the SM fermions has the form 
\bea
{\cal{L}} &=& \frac{g}{\sqrt{2}}V_{ f^\prime f} \bar{f}^\prime \gamma^\mu( g^{f^\prime f}_L P_L +  g^{f^\prime f}_R P_R) f W^\prime_\mu + ~h.c.,
\label{wprime}
\eea
where  $f^\prime$ and $f$ refer to the fermions and $g^{f^\prime f}_{L,R}$ are the left- and the right-handed couplings of the $W^\prime$. {}For a SM-like $W^\prime$ boson, $g^{f^\prime f}_{L}=1 $ and $g^{f^\prime f}_{R}= 0 $. We will assume $g^{f^\prime f}_{L,R}$ to be real. Constraints on the couplings in Eq.~(\ref{wprime}) come from the hadronic $\tau$ decays. We will consider constraints from the decays $\taud$ and $\tauv$.

The branching ratio for  $\tau^- \to \pi^- \nu_\tau$   is
\begin{eqnarray}
Br^{SM+ W^\prime}_{\tau^- \to \pi^- \nu_\tau} &=&  Br^{SM}_{\tau^- \to \pi^- \nu_\tau}  (1+ r^\pi_{W^\prime})^2\,,
\end{eqnarray}
where the  $W^\prime$  contribution is  
\begin{eqnarray}
r^\pi_{W^\prime}  &=&  x_{W^\prime} g^{\tau \nu}_L (g^{ud}_L-g^{ud}_R)\,,
\label{rpi}
\end{eqnarray}
and $x_{W^\prime} = m^2_W/M^2_{W^\prime} $. The branching ratio for the $\tau^- \to \rho^- \nu_\tau$  process is
\begin{eqnarray}
Br^{SM+ W^\prime}_{\tau^- \to \rho^- \nu_\tau} &=&  Br^{SM}_{\tau^- \to \rho^- \nu_\tau}  (1+ r^\rho_{W^\prime})^2\,,
\end{eqnarray}
with the $W^\prime$  contribution  
\begin{eqnarray}
r^\rho_{W^\prime}  &=&  x_{W^\prime} g^{\tau \nu}_L (g^{ud}_L+g^{ud}_R)\,.
\label{rrho}
\end{eqnarray}
The SM  branching ratio is related  to the decay rate as $ Br^{SM}_{\tau^- \to \rho^- \nu_\tau} = \Gamma^{SM}_{\tau^- \to \rho^- \nu_\tau}/\Gamma_\tau$ with 

\begin{eqnarray}
\Gamma^{SM}_{\tau^- \to \rho^- \nu_\tau} &=&  \frac{G^2_F}{16 \pi} |V_{ud}|^2 f^2_\rho m^3_\tau \Big(1-\frac{m^2_\rho}{m^2_\tau}\Big)^2 \Big(1 + \frac{2 m^2_\rho}{m^2_\tau}\Big)\, ,
\end{eqnarray}
where $f_\rho = 223$ MeV \cite{Wang:2006un}. Further, the SM branching ratio  can also be expressed as \cite{barish}
\begin{eqnarray}
Br^{SM}_{\tau^- \to \rho^- \nu_\tau}  &=&  1.23 Br(\tau^- \to \nu_\tau e^- \bar{\nu}_e) = 21.92 \pm 0.05 \%\,.
\end{eqnarray}
The measured branching ratio is $Br(\tau^- \to \rho^- \nu_\tau)_{exp} = (23.1 \pm 0.98)\%$ \cite{pdg}.  

\begin{figure}[htb!]
\centering
 \includegraphics[width=7.25cm]{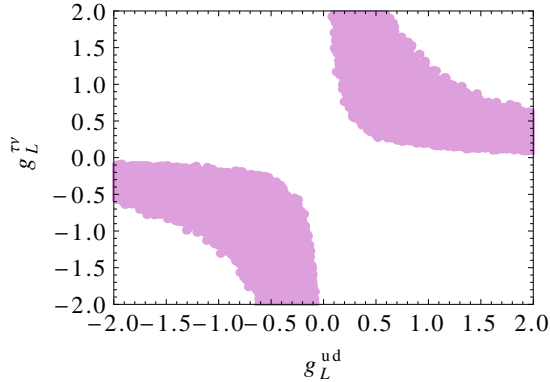}~~~
\caption{The constraints on the $W^\prime$ couplings without right-handed coupling at $M_{W^\prime}=500-1000$ GeV. The constraints are from  $\tau^- \to \pi^- \nu_\tau$ and $\tau^- \to \rho^- \nu_\tau$  branching ratios. The errors in the branching ratios are varied within $1\sigma$. The colored regions are allowed. }
\label{couplWpplotL}
\end{figure}

\begin{figure}[htb!]
\centering
 \includegraphics[width=5.25cm]{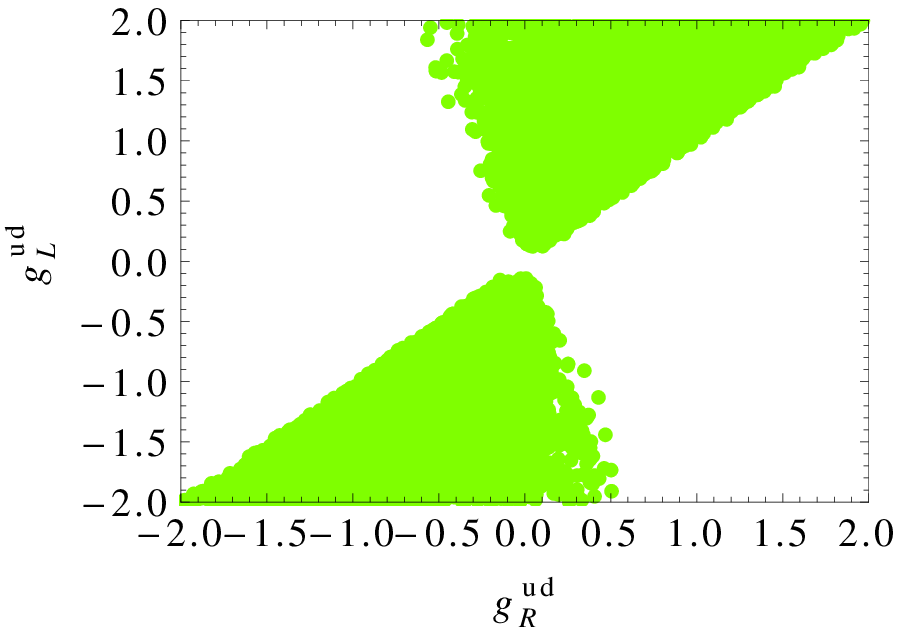}~
 \includegraphics[width=5.25cm]{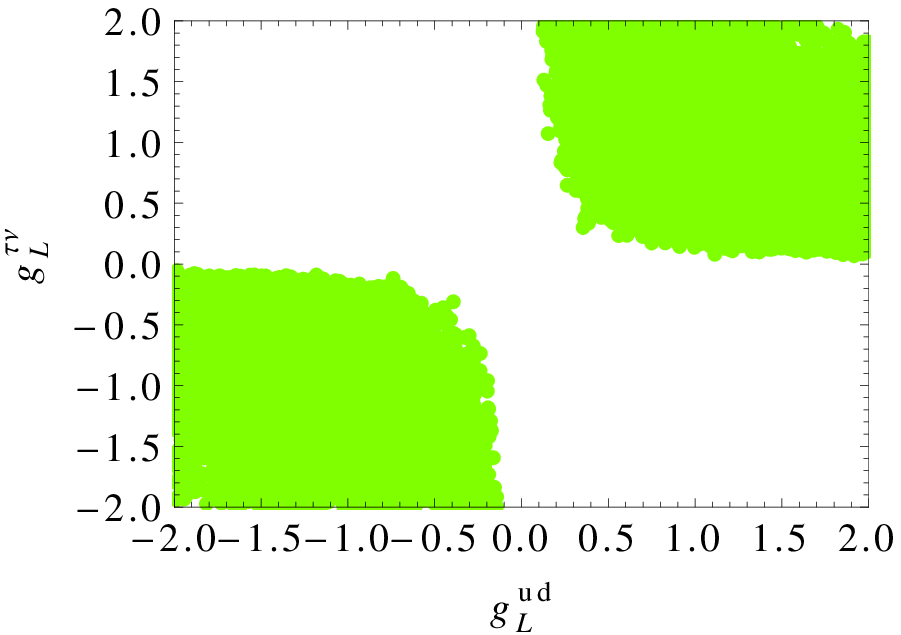}~
 \includegraphics[width=5.25cm]{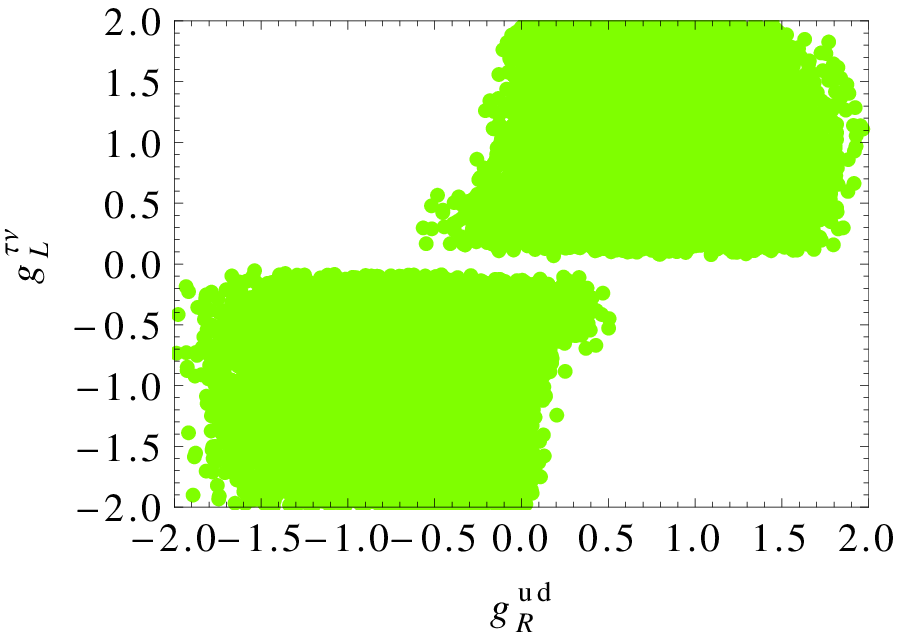}~
\caption{The constraints on the $W^\prime$ couplings with both left- and right-handed couplings at $M_{W^\prime}=500-1000$ GeV. The constraints are from  $\tau^- \to \pi^- \nu_\tau$ and $\tau^- \to \rho^- \nu_\tau$  branching ratios. The errors in the branching ratios are varied within $1\sigma$. The colored regions are allowed.}
\label{couplWpplotLR}
\end{figure}

Figures \ref{couplWpplotL} and \ref{couplWpplotLR} show the allowed regions for the $W^\prime$ couplings. The couplings are uniformly varied in the range $[-2,2]$ and constrained by the measured $\tau^- \to \pi^- \nu_\tau$ and $\tau^- \to \rho^- \nu_\tau$ branching ratios with $1\sigma$ errors. From Eqs.~(\ref{rpi}, \ref{rrho}), the  case with a pure left-handed $W^{\prime}$ coupling  is allowed, as shown in Figs.~(\ref{couplWpplotL}, \ref{couplWpplotLR}). The constraints on the effective operator are consistent with the one in Ref. \cite{biggio}. From Eq.~\ref{wprime} the NSI parameter 
$\varepsilon_{\tau\tau}^{ud(L,R)}$  defined in  Ref.~\cite{biggio} is given as $\varepsilon_{\tau\tau}^{ud(L,R)} \equiv g_L^{\tau \nu} g_{(L,R)}^{ud}( \frac{M_W}{M_{W'}})^2 $.

In the presence of the $W^\prime$ gauge boson, we can obtain the modified differential cross section for the reaction $\scat$ as
\bea
\label{Wprimediff}
\frac{d\sigma_{SM + W^\prime}(\nu_\tau)}{dt} = \frac{M^2 G^2_F \cos^2{\theta_c}}{8 \pi E^2_\nu }   \Big[A^\prime + B^\prime \frac{(s-u)}{M^2}  +C^\prime \frac{(s-u)^2}{M^4}   \Big],
\eea
where the coefficients $ A^\prime, B^\prime, C^\prime  $ include both the SM and $W^\prime$ contributions.  The expressions for these  coefficients  are given in the Appendix. 

\begin{figure}[htb!]
\centering
 \includegraphics[width=7.25cm]{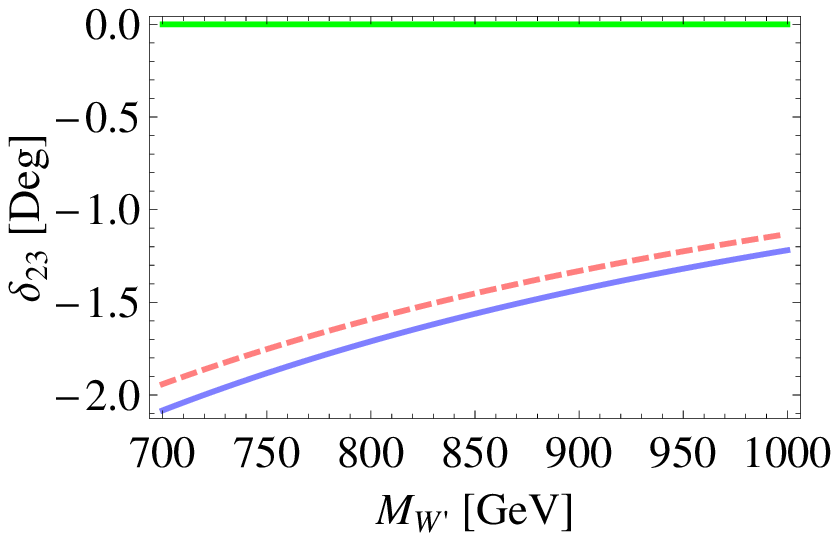}~~~
 \includegraphics[width=7.25cm]{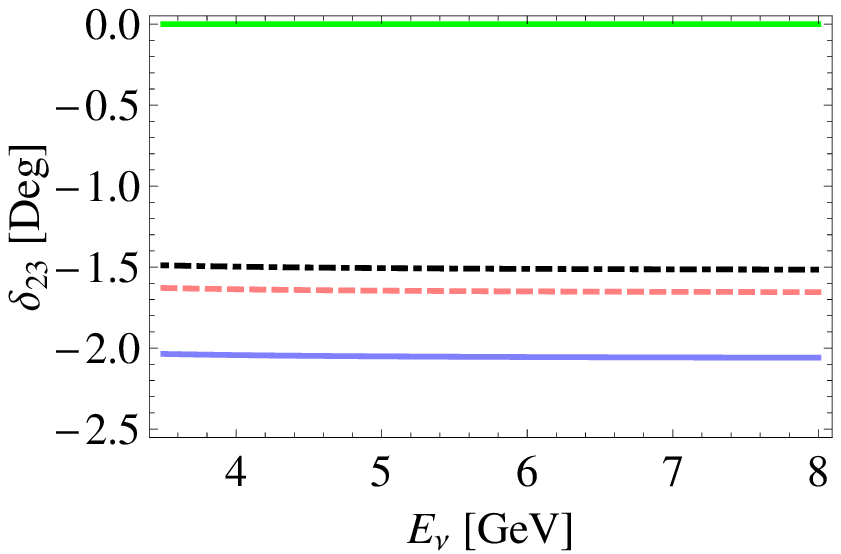}~~~
\caption{ The left (right) panel illustrates the deviation $\delta_{23}$ with the $W^\prime$ mass ($E_\nu$) when only left-handed $W^\prime$ couplings are present. The lines show predictions for some representative values of the $W^\prime$ couplings $(g^{\tau \nu_\tau}_L, g^{ud}_L)$  taken from Fig.~(\ref{couplWpplotL}). The green line corresponds to the SM prediction. The blue (solid, lower) line in the left  figure corresponds to (0.69, 0.89)  at $E_{\nu} = 5$  GeV, and the blue (solid, lower) line in the right figure corresponds to (1.42, 0.22) at $ M_{W^\prime} = 500 $ GeV. Here, we use the best-fit value $  \theta_{23} = 42.8^\circ$ \cite{GonzalezGarcia:2010er}.}
\label{delt23vsMwponlygL}
\end{figure}

For a SM-like $W^\prime$ boson, with right-handed couplings ignored, the structure of the differential cross section is similar to the one in the SM case. Hence, the $W^\prime$ contribution relative to the SM $r_{W^\prime}^{23} = \frac{\sigma_{W^\prime}(\nu_\tau)}{\sigma_{SM}(\nu_\tau)}$ does not depend on the incident neutrino energy $E_\nu$. We find  $r_{W^\prime}^{23} \sim 5\%$  at $M_{W^\prime} = 500$  GeV from the hadronic tau decay constraints in Fig.~(\ref{couplWpplotL}).  The variation of  $\delta_{23}$ with the  $W^\prime$  mass  is shown in Fig.~(\ref{delt23vsMwponlygL}). In this case, $\delta_{23}$ is always negative  and can reach up  to $-5^\circ$  at $M_{W^\prime} = 500$  GeV. Note that $\delta_{23}$  does not dependent on $E_\nu$ either.

\begin{figure}[htb!]
\centering
 \includegraphics[width=7.25cm]{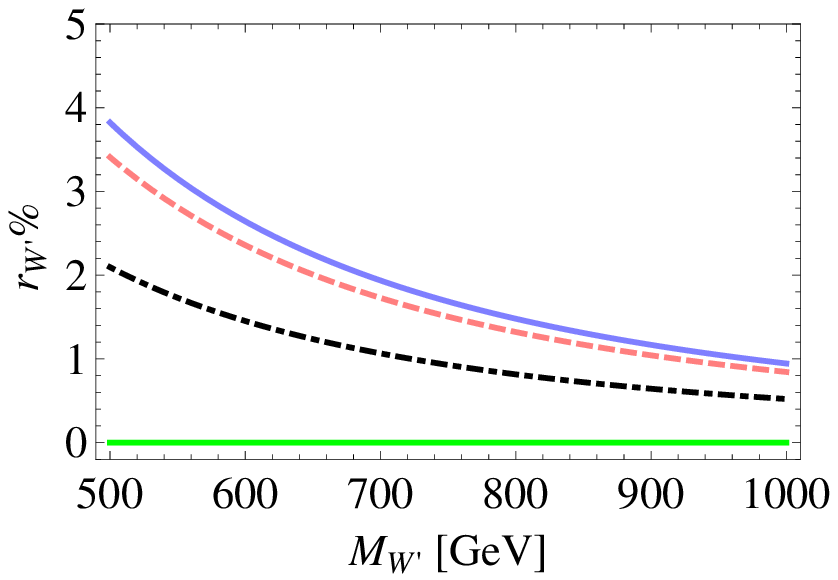}~~~
  \includegraphics[width=7.25cm]{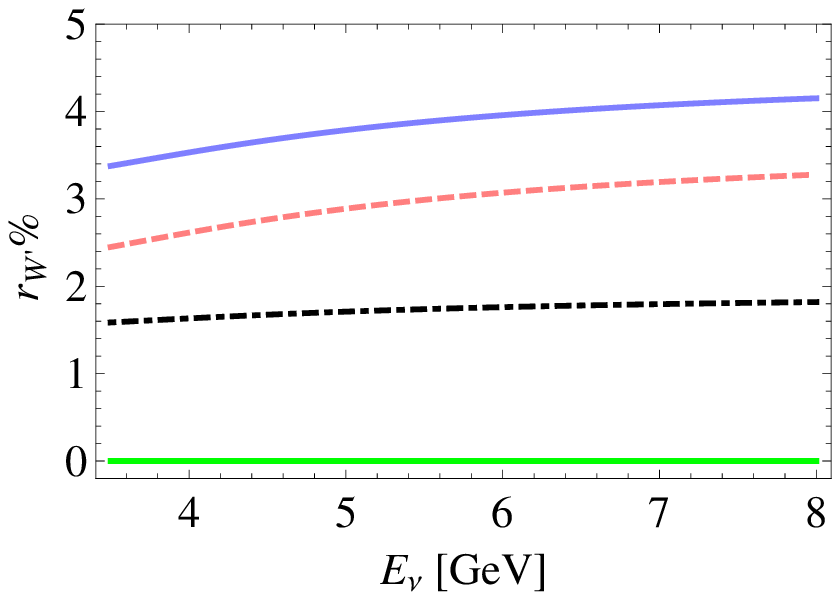}~~~
\caption{The left (right) panel illustrates the variation of $r_{W'}^{23}\%$ in 
$\scat$ scattering with the $W^\prime$ mass ($E_\nu$) when both left- and right-handed  $W^\prime$ couplings are present. The lines show predictions for some representative values of the $W^\prime$ couplings $(g^{\tau \nu_\tau}_L, g^{ud}_L, g^{ud}_R)$  taken from Fig.~(\ref{couplWpplotLR}). The green line corresponds to the SM prediction. The blue (solid, upper) line in the left  figure corresponds to (-0.94 ,  -1.13 , -0.85)  at $E_{\nu} = 5$  GeV, and the blue (solid, upper) line in the right  figure corresponds to (1.23 , 0.84 , 0.61) at $ M_{W^\prime} = 500 $ GeV. Here, we use the best-fit value $  \theta_{23} = 42.8^\circ$ \cite{GonzalezGarcia:2010er}. }
\label{dsigplWp}
\end{figure}

\begin{figure}[htb!]
\centering
 \includegraphics[width=7cm]{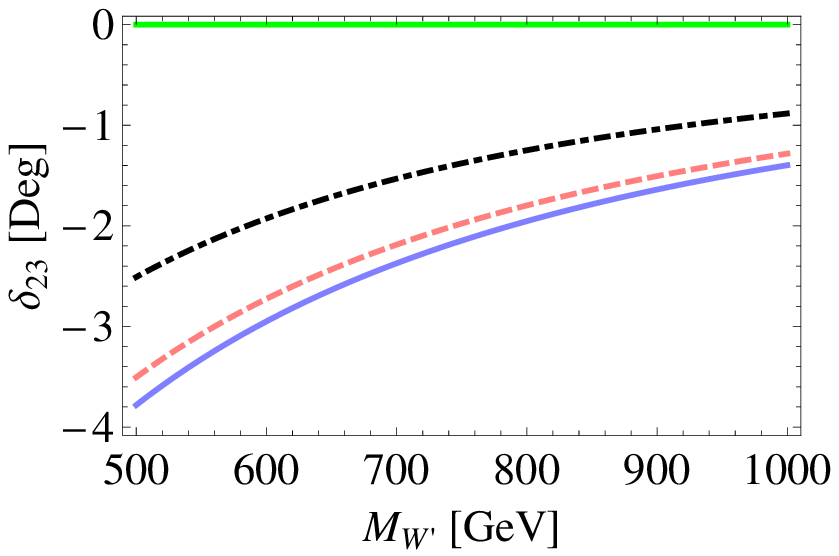}~
  \includegraphics[width=7cm]{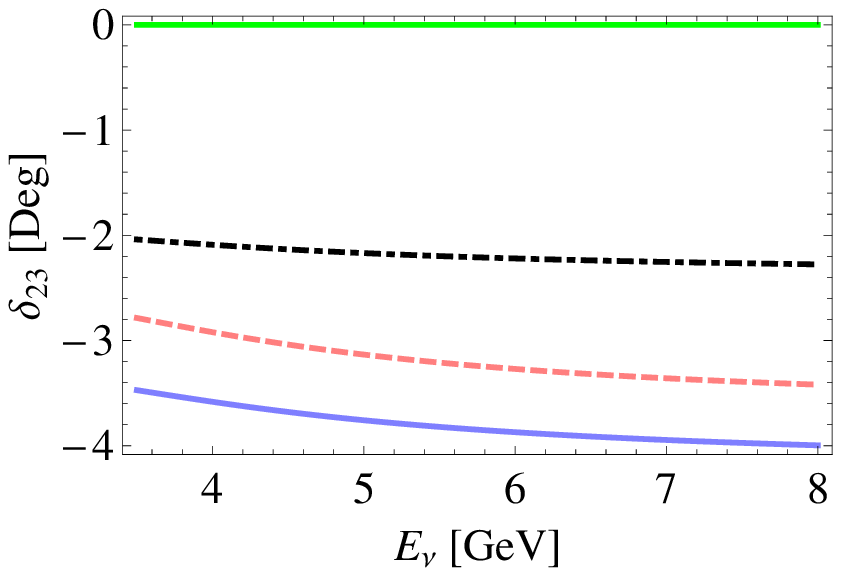}~
\caption{ The left (right) panel illustrates the deviation $\delta_{23}$ with the $W^\prime$ mass ($E_\nu$) when both the left- and right-handed  $W^\prime$ couplings are present. The lines show predictions for some representative values of the $W^\prime$ couplings $(g^{\tau \nu_\tau}_L, g^{ud}_L, g^{ud}_R)$  taken from Fig.~(\ref{couplWpplotLR}). The green line corresponds to the SM prediction. The blue (solid, lower) line in the left  figure corresponds to (-0.94 ,  -1.13 , -0.85)  at $E_{\nu} = 5$  GeV, and the blue (solid, lower) line in the right  figure corresponds to (1.23 , 0.84 , 0.61) at $ M_{W^\prime} = 500 $ GeV. Here, we use the best-fit value $  \theta_{23} = 42.8^\circ$ \cite{GonzalezGarcia:2010er}.}
\label{delt23MwpgLgR}
\end{figure}

Next, we consider  the right-handed couplings also. The  variation of $r_{W^\prime}^{23} \%$ with $M_{W^\prime}$ in this case is shown in Fig.~(\ref{dsigplWp}). The $r_{W^\prime}^{23} \%$ values are mostly positive which, in turn, leads to $\delta_{23}$ being mostly negative.  We find that $r_{W^\prime}^{23} \%$   depends slightly on the neutrino energy. The variation of  $\delta_{23}$ with the  $W^\prime$  mass and $E_\nu$  are  shown in Fig.~(\ref{delt23MwpgLgR}). 

The  $W^\prime$ contribution to the interaction $\scatanti$ leads to the following differential cross section:
\bea
\label{Wprimediff000}
\frac{d\sigma_{SM + W^\prime}(\bar{\nu}_\tau)}{dt} = \frac{M^2 G^2_F \cos^2{\theta_c}}{8 \pi E^2_\nu }   \Big[A^\prime - B^\prime \frac{(s-u)}{M^2}  +C^\prime \frac{(s-u)^2}{M^4}   \Big].
\eea
The differential cross section of the antineutrino scattering is relatively smaller than the corresponding one for the neutrino scattering because of the negative sign of the $B$ coefficient. Thus, the value of the ratio $r_{W^\prime}^{13} = \frac{\sigma_{W^\prime}(\bar{\nu}_\tau)}{\sigma_{SM}(\bar{\nu}_\tau)}$ is  smaller than the corresponding ratio, $r_{W^\prime}^{23}$. Because of the smallness of $\theta_{13}$ and $r_{W^\prime}^{13}$, the NP effect on  the extraction of $\theta_{13}$ is  small. Achieving large $r_{W^\prime}^{13}$ within the constraints given in Fig.~\ref{couplWpplotLR} is difficult in this model. This means the effect of the NP contribution in $\delta_{13}$ is very small and we do not plot the results of this calculations.

\section{Conclusion}
In this paper we calculated the effect of a charged Higgs and a $W^\prime$ contribution to  $\scat$ and $\scatanti$ scattering. We constrained the parameters of both the models from $\taud$ and $\tauv$ decays. Corrections to the SM contribution to $\scat$ and $\scatanti$ impact the extraction of the neutrino atmospheric mixing angles $\theta_{23}$ and $\theta_{13}$.  
We found that the charged Higgs model can produce significant corrections to $\delta_{23, 13} $ that measure the deviation of the actual $\theta_{23, 13} $ from the $(\theta_{23, 13})_{SM}$ angles which are extracted assuming the SM $\nu_\tau / \bar{\nu}_\tau$ scattering cross sections. The $W'$ model effect generates a large deviation $\delta_{23}$ but negligibly small $\delta_{13}$.
As $\theta_{13}$ is smaller than $\theta_{23}$ larger NP in
$\scat$ and $\scatanti$  is required to
produce  effects in $ \delta_{13}$ similar in size to $\delta_{23}$.

When a charged Higgs is involved, $\delta_{23, 13}$ are negative. This is because there is no interference of the charged Higgs contribution with the SM contribution, for massless neutrinos, and so the cross sections for $\scat$ and $\scatanti$ are always larger than the SM cross sections. This means that experiments should measure $\theta_{23, 13}$ larger than the present values in the presence of a charged Higgs contribution. In this case we also found that $\delta_{23, 13}$ increase in magnitude with the neutrino energy. Hence, a possible sign of the charged Higgs effect would be a measurement of $\theta_{23, 13}$ that shows an increase with increasing neutrino energy. 

For the $W^\prime$ model we calculated a significant contribution to $\delta_{23}$ which can be both positive and negative, but is mostly negative. The deviation $\delta_{23}$  was found to be independent of the neutrino energy for a left-handed $W^\prime$ but neutrino energy dependent when both left- and right-handed $W^\prime$ chiralities were present. A negligibly small deviation, $\delta_{13}$, was found in the $W'$ model because of the small value of $\theta_{13}$. 


We have presented in this paper a first estimation of the charged Higgs and $W^\prime$ effects in the extraction of $\theta_{23}$ and $\theta_{13}$. We hope more detailed calculations including nuclear  as well as detector effects, will be done to find out whether these new physics effects can be observed at present $\nu_{\tau}/\bar{\nu}_{\tau}$ appearance experiments and/or to motivate new experiments that can detect these effects.

\section*{Acknowledgements} We thank Sandip Pakvasa and Nita Sinha for useful discussions.
This work was financially supported  
in part by the National Science
Foundation under Grant No.\ NSF PHY-1068052.

\appendix

\section*{\centering Appendix}

\subsection*{Hadronic form factors}

The expressions for the vector and axial-vector hadronic currents in Eq.~\ref{hadcureq000} are
\bea
\label{vecAxME}
\bra p(p^\prime)|V_\mu| n(p) \ket & =& \bar{u}_p(p^\prime) \Big[\gamma_\mu F^V_1  + \frac{i}{2 M} \sigma_{\mu \nu} q^\nu F^V_2   + \frac{q_\mu}{M} F_S  \Big] u_n(p),\nonumber\\
-\bra p(p^\prime)|A_\mu | n(p) \ket   &=& \bar{u}_p(p^\prime) \Big[\gamma_\mu  F_A  + \frac{i}{2 M} \sigma_{\mu \nu} q^\nu  F_T  + \frac{q_\mu}{ M}  F_P \Big]\gamma_5 u_n(p).
\eea
Here $q = p^\prime-p$ and the form factors $F_i$ are functions of $t=q^2$.  The parametrizations of the  axial-vector and pseudoscalar form factors are \cite{SM-calc}
\bea
\label{FFAxial}
F_{A}(t) &=& F_A(0) \Big(1- \frac{t}{M^2_A}\Big)^{-2}, \nn \\
F_{P}(t) &=&  \frac{2 M^2 F_{A}(0)}{m^2_\pi - t} \,,
\eea
where $F_A(0)= - 1.2695$ is the axial coupling \cite{pdg}, $m_\pi$ is the  charged pion mass, and $M_A = 1.35$ GeV is the axial-vector mass \cite{Axial mass}. The  expression  for $F_{P}(t)$ can be shown to be true at low energy, where the predictions of chiral perturbation theory are valid \cite{fearing}. We have assumed the relation to hold at high $t$ also. Note that $F_{A}(0)$ is sometimes replaced by $F_{A}(t)$, which gives similar results for $F_{P}(t)$ at low $t$ but very different results at high $t$.

 The Dirac and Pauli form factors $F^V_{1,2}$ are 
\bea
\label{FF12V}
F^V_{1}(t) &=& \frac{G_E (t) - x_t G_M (t)}{1- x_t},~
F^V_{2}(t) = \frac{G_M (t) - G_E (t)}{1- x_t} \,,
\eea
where $x_t = t/4 M^2$ and
\bea
\label{GEGM}
G_M = G^p_M- G^n_M,~G_E = G^p_E- G^n_E\,.
\eea
Here $G^{p,n}_E$ and $G^{p,n}_M$ are the electric and magnetic form factors of the proton and neutron, respectively. The simplest parametrizations of these form factors are given by the dipole approximation
\bea
\label{dipoleapp}
 G^p_E \approx G_D,~ G^n_E \approx 0,~  G^p_M \approx \mu_p G_D,~ G^n_M \approx \mu_n G_D\,,
\eea
where $G_D = (1 - t/M^2_V)^{-2}$, $M_V = 0.843$ GeV is the vector mass, and $\mu_p(\mu_n)$  is the anomalous  magnetic moment of the proton (neutron) \cite{Kuzmin:2007kr}. 

In the presence of the charged Higgs, applying the equation of motion to the  hadronic matrix elements for the scalar and pseudoscalar currents for the process $\scat$ gives
\beq
\bra p(p^\prime)|\bar{u} (g_S^{u_i d_j}- g_P^{u_i d_j} \gamma_5) d|n(p) \ket = V_{ud}\; \bar{p}(p_4) \left(g_S^{u_i d_j} G_S + g_P^{u_i d_j} G_P \gamma_5 \right) n(p_2),
\eeq
or
\bea
\label{vecSca}
\bra p(p^\prime)|\bar{u} d|n(p) \ket &=&  \bar{p}(p_4) G_S n(p_2), \nn \\
- \bra p(p^\prime))|\bar{u} \gamma_5 d|n(p) \ket &=&  \bar{p}(p_4) G_P\gamma_5 n(p_2) \,,
\eea
where 
\bea
G_S (t) &=&  r_N  F^V_1(t),\;\; \mbox{with}\;\;  r_N  =\frac{M_n-M_p}{(m_d-m_u)}\sim  {\cal{O}} (1), \nonumber\\
G_P (t)  &=&  \frac{M[ F_A(t) + 2 x_t  F_P(t)]}{\bar{m}_q},
\eea
with $\bar{m}_q =(m_u+m_d)/2$. 
In the $W'$ model, the current has both $V\pm A$ structures. One has to calculate the matrix element,
\bea
 \label{hadcureq000}
\bra p(p^\prime)|J^+_\mu |n(p) \ket &=&  V_{ud}\; \bra p(p^\prime)|\bar{u}\left( g_L^{ud}\gamma_\mu(1 -\gamma_5)+g_R^{ud}\gamma_\mu(1 +\gamma_5)\right) d| n(p) \ket.
 \eea


\subsection*{Kinematic details}

The Mandelstam variables in terms of $E_\nu$ and the lepton energy $E_l$ are
\bea
\label{Manvar}
s &= & M^2 + 2 M E_\nu,~ t = 2 M(E_l -E_\nu),  \nn\\  s-u &=& 4 M E_\nu + t -m^2_l.
\eea
 Then $t$ and $E_l$ lie in the intervals 
 \bea
 \label{tint}
  m^2_l- 2 E_\nu^{\rm cm} \left( E_l^{\rm cm} + p_l^{\rm cm} \right)   \leq  t \leq  
  m^2_l- 2 E_\nu^{\rm cm} \left( E_l^{\rm cm} - p_l^{\rm cm}\right) ,
 \eea
 \bea
 E_\nu + \frac{m_l^2-2E_\nu^{\rm cm} (E_l^{\rm cm} + p_l^{\rm cm})}{2M} \leq  E_l \leq E_\nu + \frac{m_l^2-2E_\nu^{\rm cm} (E_l^{\rm cm} - p_l^{\rm cm})}{2M},
 \eea
where the energy  and momentum of the lepton and the neutrino in the center of mass (cm) system are
\bea
E_\nu^{cm} &=& \frac{(s-M^2)}{2 \sqrt{s}},\;\; p_l^{\rm cm} = \sqrt{(E_l^{cm})^2-m^2_l},\nonumber\\
E_l^{cm} &=& \frac{(s-M^2+m^2_l)}{2 \sqrt{s}}.
\eea
The threshold neutrino energy to create the charged lepton partner is given by
\beq
E_{\nu_l}^{\rm th} = \frac{(m_l + M_p)^2 - M_n^2}{2 M_n},
\eeq 
 where $m_l,\; M_p,\; M_n$ are the masses of the charged lepton, proton, and neutron, respectively. In our case, the threshold energy of the tau neutrino is $E_{\nu_\tau}^{\rm th} = 3.45$ GeV.

The differential cross section in the laboratory frame is given by
\beq
\frac{d\sigma_{tot}(\nu_l)}{dt} = \frac{|\bar{\mathcal{M}}|^2 }{64 \pi E^2_\nu M^2 }.
\eeq
\vspace*{0.2cm}


The expressions for the coefficients $f_{SM}\; (f = A, B, C)$  in the SM differential cross section  [see Eq.(\ref{SMdiff})] are 
\bea
\label{SMdiffco}
A_{SM} &=& 4 (x_t - x_l) \Big[(F_1^{V})^2 (1+x_l+x_t)+ (F_A)^2(-1 + x_l + x_t) + (F_2^{V})^2 (x_l+x_t^2+x_t) \nn \\
& &  + 4 F_P^2 x_l x_t + 2 F_1^{V} F_2^{V} (x_l+2 x_t) + 4 F_A F_P x_l\Big]\,, \nn  \\
B_{SM} &=&  4  x_t   F_A (F_1^V + F^V_2) \,, \nn \\
C_{SM} &=& \frac{(F_1^V)^2 + F_A^2 -x_t (F_2^V)^2 }{4} \,,
\eea
where $x_l = m^2_l /4 M^2$. \vspace*{0.2cm}

The expressions for the quantities $A^{I,P}_{H}$  and $B^{I}_{H}$ in the  differential cross section in Eq.~(\ref{Higgsdiff}) are

\bea
\label{Phiecoeff}
A^{I}_{H} &=& 2 \sqrt{ x_l} (x_t - x_l) g_P^{ud}  (g_S^{l\nu_l} - g_P^{l\nu_l}) G_P (F_A + 2 F_P x_t)\,,\nn\\ 
B^{I}_{H} &=& \frac{1}{2} \sqrt{ x_l} g_S^{ud} (g_S^{l\nu_l} - g_P^{l\nu_l})G_S (F_1^V + F^V_2 x_t)\,,\nn\\ 
A^{P}_{H} &=& 2 (x_t - x_l)(|g_S^{l\nu_l}|^2 +    |g_P^{l\nu_l}|^2) (|g_P^{ud}|^2 G_P^2 x_t + |g_S^{ud}|^2 G_S^2 (x_t - 1))\,.
\eea
For the 2HDM II model couplings, $g_{S, P}$ are given in Eq.~(\ref{2HDMcoup}). Note that the interference terms $A^{I}_{H}$ and $B^{I}_{H}$ vanish in this model.\vspace*{0.2cm}

The expressions for the quantities $f^{\prime}\; (f = A, B, C)$ in the  differential cross section in  Eq. (\ref{Wprimediff})  are 

\bea
\label{Wprimecoeff}
A^\prime &=& 4 (x_t - x_l) \Big[(1 + r^\rho_{W^\prime})^2 \Big((F_1^{V})^2 (1+x_l+x_t) +   2 F_1^{V} F_2^{V} (x_l+2 x_t) + (F_2^{V})^2 (x_l+x_t^2+x_t) \Big) \nl && + (1 + r^\pi_{W^\prime})^2 \Big((F_A)^2(-1 + x_l + x_t) + 4 F_A F_P x_l +4 F_P^2 x_l x_t\Big)\Big]\,, \nl 
B^\prime &=& 4 Re[(1 + r^\rho_{W^\prime}) (1 + r^{\pi*}_{W^\prime})] x_t   F_A (F_1^V + F^V_2)\,,\nl
C^\prime &=& \frac{1}{4} \Big[ (1 + r^\rho_{W^\prime})^2 ((F_1^V)^2 -x_t (F_2^V)^2) + (1 + r^\pi_{W^\prime})^2  F_A^2 \Big]\,.
\eea
In the absence of $W^\prime$ contributions, the $f^{\prime}$'s  reduce to the respective  SM results in Eq.~\ref{SMdiffco}. 







\end{document}